\documentclass[twocolumn,showpacs,pre]{revtex4}
 \usepackage{graphicx,amsmath}
\begin{document}

\title{Probabilistic Description of Traffic Breakdowns}

\author{Reinhart K\"{u}hne}
\affiliation{German Aerospace Center, Institute of Transport Research,
Rutherfordstra{\ss}e 2, 12489 Berlin, Germany}

\author{Reinhard Mahnke}
\affiliation{Fachbereich Physik, Universit\"{a}t Rostock, D--18051 Rostock,
Germany}

\author{Ihor Lubashevsky}
\affiliation{Theory Department, General Physics Institute, Russian Academy of
Sciences, Vavilov str., 38, Moscow, 119991, Russia}

\author{Jevgenijs Kaupu\v{z}s}
\affiliation{Institute of Mathematics and Computer Science, University of
Latvia, 29 Rainja Boulevard, LV--1459 Riga, Latvia}

\date{\today}

\begin{abstract}
We analyze the characteristic features of traffic breakdown. To describe this
phenomenon we apply to the probabilistic model regarding the jam emergence as
the formation of a large car cluster on highway. In these terms the breakdown
occurs through the formation of a certain critical nucleus in the metastable
vehicle flow, which enables us to confine ourselves to one cluster model. We
assume that, first, the growth of the car cluster is governed by attachment of
cars to the cluster whose rate is mainly determined by the mean headway
distance between the car in the vehicle flow and, may be, also by the headway
distance in the cluster. Second, the cluster dissolution is determined by the
car escape from the cluster whose rate depends on the cluster size directly.
The latter is justified using the available experimental data for the
correlation properties of the synchronized mode. We write the appropriate
master equation converted then into the Fokker-Plank equation for the cluster
distribution function and analyze the formation of the critical car cluster due
to the climb over a certain potential barrier. The further cluster growth
irreversibly gives rise to the jam formation. Numerical estimates of the
obtained characteristics and the experimental data of the traffic breakdown are
compared. In particular, we draw a conclusion that the characteristic intrinsic
time scale of the breakdown phenomenon should be about one minute and explain
the case why the traffic volume interval inside which traffic breakdown is
observed is sufficiently wide.
\end{abstract}
\maketitle

\section{Introduction. Traffic breakdown as the nucleation phenomenon}

The spontaneous formation of traffic jams on highways has attracted attention
for the last years because of two reasons. The former is the importance of this
problem for traffic engineering especially concerning the feasibility of
attaining the limit capacity of traffic networks and quantifying it. The latter
is a fact that the jam formation nicely exemplifies the existence of various
phase states and transitions between them in statistical systems comprising
elements with motivated behavior, which is a novel branch of physics. According
to the modern notion proposed by Kerner (see, e.g. Ref.~\onlinecite{K1,K1a,K2})
based on the experimental data \cite{K3,K4,K5,K6} the jam formation is of
sufficiently complex nature. In particular, it proceeds mainly through the
sequence of two phase transitions: free flow (F) $\rightarrow $ synchronized
mode (S) $\rightarrow $ stop-and-go pattern (J) \cite{K6}. Both of these
transitions are of the first order, i.e. they exhibit breakdown, hysteresis,
and nucleation effects \cite{K5}. The F$\,\rightarrow\,$J transition can occur
directly if only the synchronized mode is suppressed by a road heterogeneity
\cite{K2}. The recent analysis of single-vehicle data by Neubert \textit{et
al.} \cite{N1}, in particular, conformed these features and also discovered
fundamental microscopic properties distinguishing the synchronized mode from
the other traffic states.

Theoretical description of the jam formation is far from being developed well
because the synchronized mode possesses extremely complex structure \cite{K4}.
For example, it comprises a certain continuum of quasistable states, so,
matches a whole two-dimensional region on the phase plane ``vehicle
density~--~traffic volume'' in contrast to the free flow state. However,
tackling the question of how to regulate traffic flow on highways, for example,
by controlling the speed limitation in order to prevent the jam formation we
may rough out the problem. Indeed, for this purpose it is sufficient to analyze
the conditions giving rise to jams rather then the jam evolution itself. Such a
standpoint is justified, in part, by the aforementioned phase transitions being
of the first order.
The free flow state, presumably, should have the
feasibility to exist at the given car density or inside its certain
neighborhood. This might be the necessary requirement for jam formation at a
fixed vehicle density, or for appearance of both the jam phase and the
synchronized mode at a fixed traffic volume.
Second, the jam formation proceeds via the nucleation mechanism but not in a
regular manner. Therefore, the key point in the emergence of a jam is the
random occurrence of its critical nucleus inside the synchronized mode or free
flow.

The jam formation manifests itself in the traffic breakdown, i.e. in a sharp
drop of the traffic volume to a substantially lower value. Detecting these
events one can get the rate of the critical nucleus generation depending on the
road conditions and the traffic state. In this way the main attention is
shifted to the experimental and theoretical analysis of the probabilistic
features of jam formation regarding the characteristic mean values of the
traffic volume as phenomenological parameters \cite{Br1,Br2,Br3,Br4}. Such a
probabilistic description of the traffic breakdown is the main purpose of the
present paper.

At the first glance the problem seems hopeless until the model of the
synchronized mode is developed. Nevertheless, there are circumstances enabling
us to make a step towards this problem right now (see also
Ref.~\onlinecite{KBr}). The matter is that the F\,$\rightarrow$\,S transition
is of another nature than the S\,$\rightarrow$\,J transition. The former is due
to a sharp decrease in the overtaking frequency, giving rise to the
synchronized mode, whereas the latter is caused by the pinch effect (see, e.g.
Ref.~\onlinecite{K1,K1a,K2}). Thereby the main case of the F\,$\rightarrow$\,S
transition is not fluctuations in the vehicle density and velocity but in other
characteristic parameters of the traffic flow (cf. also Ref.~\onlinecite{W1}).
By contrast, just these fluctuations give rise to the jam emergence in the
synchronized traffic flow. As a result, the threshold of the F$\,\rightarrow
\,$S transition turns out to be remarkably less than that of the jam formation
and attained at lower values of the vehicle density. So, the generation rate of
critical nuclei for the former transition has to be great in comparison with
the latter one. Thus, on time scales characterizing the occurrence of the jam
critical nuclei the traffic state with respect to the transitions between the
free flow and synchronized mode is quasistationary. Therefore the formation of
a jam critical nucleus is the leading nonequilibrium process limiting the
traffic breakdown. The latter feature allows us to confine our consideration
solely to the jam nucleus generation and to regard the synchronized mode and
the free flow phase (if they coexist in the case under consideration) as one
traffic state. Moreover, since a jam forms actually inside the synchronized
mode where the vehicle motion at different lane is strongly correlated we may
apply to a single-lane road approximation that treats all the cars moving at
different lanes on a multilane highway and being neighboring across the highway
as a single effective macrovehicle consisting of many cars. The macrovehicle
concept is partly justified by the empirically observed fact that fluctuations
in the downstream flow leaving a freeway bottleneck can proceed without the
traffic state change even their amplitude attains 30\% of the mean traffic
volume \cite{Br1,Br2,Br3,Br4,KBr}. In any case fluctuations in the traffic flow
volume near its breakdown are of macroscopic nature and the critical nucleus of
traffic jam has to include many vehicles. This feature is also pointed to by
the observed breakdown near a ramp-on occurring each time after a large vehicle
cluster entered the freeway stream \cite{Br1}.

It should be pointed out the real structure of congested traffic near a highway
bottleneck is sufficiently complex, it contains the region of synchronized mode
located in the close proximity of the bottleneck, the preceding upstream region
of moving narrow jams transformed into wide jams \cite{KBr}. However it is
quite reasonable to consider this structure as being induced by the traffic
breakdown processes arising inside the ``head'' of this complex jam, in the
region of synchronized mode adjacent the bottleneck. Therefore the main
characteristics of the breakdown phenomenon may be related to intrinsic
processes taking place inside the synchronized mode on not to large spatial
scales. The latter justifies our attempt to describe traffic breakdown ignoring
the complex spatial structure of the metastable traffic state inside which
critical jam nuclei originate.

Processes similar to the traffic breakdown are widely met in physical systems.
For example, water condensation in supersaturated vapor proceeds via formation
of small atom clusters of critical size. Keeping in mind this analogy between
the traffic breakdown and the phase transitions in physical systems Mahnke
\textit{et al.} \cite{M1,M2} proposed a kinetic approach based on the
stochastic master equation describing the jam formation in terms of the
attachment of individual cars to their cluster. However, the particular form of
the developed master equation does not allow for the jam formation being the
first order phase transition and, thus, the traffic breakdown. In the present
paper we generalize this kinetic approach to describe the latter phenomenon.

\begin{figure}
\begin{center}
\includegraphics[scale=1]{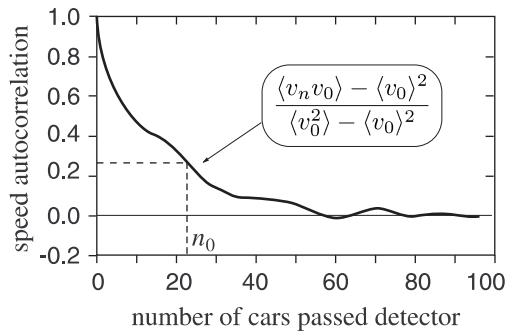}
\end{center}
\caption{Illustration of the speed autocorrelation \textsl{vs} the number of
cars passed a fixed detector that was experimentally observed by Neubert,
Santen, Schadschneider, and Schreckenberg \protect\cite{N1}.\label{SVD}}
\end{figure}

However, before passing directly to the model statement we recall the
experimental data enabling us to estimate the characteristic size $n_0$ of
vehicle clusters that are small enough so the behavior of drivers inside them
seems to be special. From our point of view the multilane correlations in the
vehicle motion are due to the drivers taking into account the behavior of all
the cars, including also the cars at the neighboring lanes, that are inside the
region accessible to observation. Therefore the synchronized mode has to
exhibit strong correlations in this region. Figure~\ref{SVD} shows the speed
autocorrelation function \textsl{vs} the number of cars passed a fixed detector
that was experimentally found in the synchronized mode \cite{N1} (see also
Ref.~\onlinecite{Hrev2}). We see that the car velocities are strongly
correlated over scales spanning some ten vehicles, i.e. a car cluster of this
size, $n_0\gtrsim 20$, makes up actually a certain fundamental unit of the
synchronized mode. In the free flow no such long-distant correlations have been
observed.

\section{Probabilistic model for the car aggregation}

\subsection{Discrete description. Master equation}

We consider traffic flow on a single-lane road and study the spontaneous
formation of a jam regarded as a large car vehicle cluster arising on the road.
Instead of dealing with a certain road part of length $L$ and imposing some
boundary conditions at its entries and exits we examine a circular road of
length $L$ with $N$ cars moving on it. All the cars are assumed to be identical
vehicles of length $l_{\text{car}}$ and can make up two phases. One of them is
the set of ``freely'' moving cars and the other is a single cluster. The
cluster is specified by its size $n$, the number of aggregated cars. Its
internal parameters, namely, the headway distance $h_{\text{clust}}$ (i.e. the
distance between the front bumper of a chosen car and the back bumper of the
following one) and, consequently, the velocity $v_{\text{clust}}$ of cars in
the cluster are treated as fixed values independent of the cluster size $n$. We
note that in the model under consideration there can be only one cluster on the
road. The ``free'' flow phase is also specified by the corresponding headway
distance $h_{\text{free}}(n)$ which, however, already depends on the car
cluster size $n$ because the larger is the cluster, the less is the number
($N-n$) of the ``freely'' moving cars and, so, the longer is the headway
distance $h_{\text{free}}(n)$.

\begin{figure}
\begin{center}
\includegraphics[scale=1]{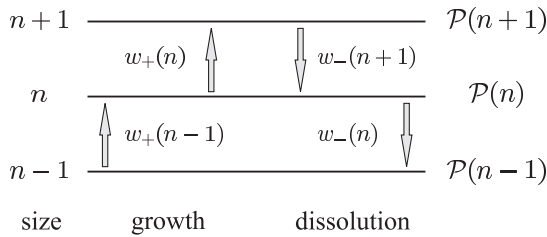}
\end{center}
\caption{Schematic illustration of the cluster transformations.\label{ME}}
\end{figure}

When a cluster arises on the road its further growth is due to the attachment
of the ``free'' cars to its upstream boundary, whereas the cars located near
its downstream boundary accelerating leave it, which decreases the cluster
size. These processes are treated as random changes of the cluster size $n$ by
$\pm 1$ (Fig.~\ref{ME}) and the cluster evolution is described in terms of time
variations of the probability function $\mathcal{P}(n,t)$ for the cluster to be
of size $n$ at time $t$. Then following Mahnke \textit{et al.} \cite{M1,M2} we
write the following master equation governing the cluster growth
\begin{align}
 \partial_t \mathcal{P}(n,t) & =
 w_{+}(n-1)\mathcal{P}(n-1,t)+w_{-}(n+1)\mathcal{P}(n+1,t)\nonumber\\
 &\hphantom{{}={}} -\left[ w_{-}(n)+w_{+}(n)\right] \mathcal{P}(n,t)\,,\label{e.1}
\end{align}
where the cluster size $n$ meets the inequality $1\leq n\leq (N-1)$ and
$w_{+}(n)$ and $w_{-}(n)$ are the transition rates illustrated in Fig.~\ref{ME}
and depending generally on the cluster size $n$. The formation and dissolution
of the maximum possible cluster containing all the cars is described by the
equation
\begin{equation}
 \partial_t \mathcal{P}(N,t)=w_{+}(N-1)\mathcal{P}
 (N-1,t)-w_{-}(N)\mathcal{P}(N,t)\,,  \label{e.2}
\end{equation}
whereas the emergence of the jam seed, the cluster consisting of one car
called below precluster, obeys the equation
\begin{equation}
\partial_t \mathcal{P}(0,t)=w_{-}(1)\mathcal{P}
(1,t)-w_{+}(0)\mathcal{P}(0,t)\,.  \label{e.3}
\end{equation}
Here the function $\mathcal{P}(0,t)$ is the probability of no cluster on the
road. At the initial time $t=0$ no cluster is assumed to be on the road:
\begin{equation}
\mathcal{P}(n,0)=\delta _{n0}\,,  \label{e.i}
\end{equation}
where $\delta _{nn^{\prime}}$ is Kronecker's symbol. The system of
equations~(\ref{e.1})--(\ref{e.3}) subject to the initial condition~(\ref{e.i})
makes up the probabilistic description of the cluster formation.

Special attention should be paid to the question as to what the precluster is.
The model proposes the following. When there is no cluster on the road, i.e.
all the cars move ``freely'' the velocity of one of them can randomly drop down
to the velocity $v_{\text{clust}}$ in the cluster. Such a car is regarded as
the precluster, a size-one cluster. When a precluster has arisen its further
evolution follows the scheme shown in Fig.~\ref{ME}. The precluster concept may
be justified by recalling the problem we deal with initially, i.e. the
breakdown processes in multilane traffic flow. The cars under consideration
actually match small vehicle clusters of the synchronized mode, macrovehicle,
arising in traffic flow on a multilane highway and comprised of real vehicles
moving synchronously at different lanes. Therefore the precluster is actually
as a certain sufficiently small cluster of the synchronized mode. Keeping in
mind the relatively low threshold of the F\,$\rightarrow$\,S transition we will
assume the precluster generation as well as the precluster dissipation to be
intensive processes so that the ``free'' flow phase, $n=0$, and the precluster
state, $n=1$, come into quasi-equilibrium on time scales needed for the
critical cluster nucleus to arise. In particular, in no case the precluster
emergence limits the cluster evolution, so, the particular details of the
precluster formation has no substantial effect on the traffic breakdown.

\begin{figure}
\begin{center}
\includegraphics[scale=0.9]{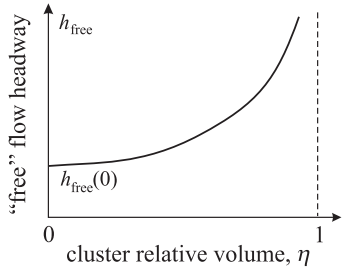}
\end{center}
\caption{\label{dvss}The headway distance $h_{\text{free}}$ in the ``free''
flow phase \textsl{vs} the cluster relative volume $\eta=n/N$. A qualitative
sketch.}
\end{figure}

At the next step we should specify the transition rates $w_{+}(n)$ and
$w_{-}(n)$. Let us apply to the optimal velocity model assuming the velocity
$v$ of the ``freely'' moving cars as well as the clustered cars to be
determined directly by the corresponding headway distance $h$ according to the
formula
\begin{equation}
v=\vartheta (h):=v_{\text{max}}\frac{h^p}{h^p+D_{\text{opt}}^{p}}\,.
\label{e.4}
\end{equation}
Here the value $D_{\text{opt}}$ is the headway distance at which drivers feel
themselves ``free'' and their velocity attains the maximum $v_{\text{max}}$.
The parameter $p>1$ allows for possible forms of the function $\vartheta(h)$,
the greater the value of $p$, the sharper the dependence $\vartheta(h)$. Case
$p=2$ is often used~\cite{M1,M2}. A car attaches itself to the cluster as fast
as the distance to the last car in the cluster decreases down to the cluster
headway $h_{\text{clust}}$, enabling us to write the following ansatz for the
attachment rate to the cluster of size $n\geq 1$
\begin{equation}
w_{+}(n)=w_{+}^{\text{ov}}(n):=\frac{\vartheta \left[ h_{\text{free}}(n)
\right] -\vartheta \left[ h_{\text{clust}}\right] }{h_{\text{free}}(n)-
h_{\text{clust}}}\,.  \label{e.5}
\end{equation}
Applying to a simple geometrical consideration and assuming $N\gg 1$ as well
as $N-n\gg 1$ we get the relationship (illustrated also in Fig.~\ref{dvss})
\begin{equation}
h_{\text{free}}(n)=h_{\text{clust}}+(l_{\text{car}}+h_{\text{clust}})\frac{\left(
\rho _{\text{lim}}-\rho\right)}{\rho (1-\eta )}\,,  \label{e.6}
\end{equation}
which together with (\ref{e.5}) gives the attachment rate as a function of the
cluster size $n$. Here we have introduced the following traffic flow
parameters: $\rho =N/L$ being the mean value of the car density on the road,
its maximum possible value $\rho
_{\text{lim}}=1/(l_{\text{car}}+h_{\text{clust}})$ for the given road, and the
relative volume $\eta =n/N$ of the cluster with respect to the initial volume
of the ``free'' flow state.

In order to compare the cluster growth due to the car attachment with the
precluster generation we specify its rate in terms of
\begin{equation}
w_{+}(0)=\epsilon w_{+}^{\text{ov}}(0)\,,  \label{e.7}
\end{equation}
where $\epsilon$ is a phenomenological factor and we formally set $n=0$ in
expression (\ref{e.5}). Keeping in mind the aforesaid about the precluster
emergence we assume the factor $\epsilon$ to be about unit, $\epsilon\sim 1$,
or at least not to be small enough to limit the cluster formation, so its
particular numerical value is of no importance.

\begin{figure}
\begin{center}
\includegraphics[scale=0.9]{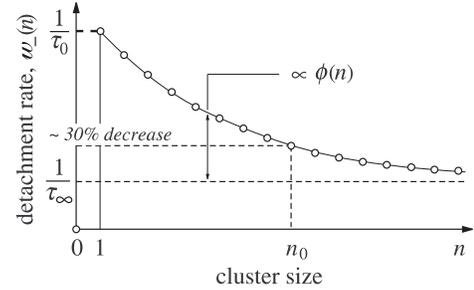}
\end{center}
\caption{The detachment rate $w_{-}(n)$ \textsl{vs} the cluster size $n$. A
qualitative sketch.\label{Fattach}}
\end{figure}

The rate of the cars escaping from the cluster at its downstream front is
written as (see also Fig.~\ref{Fattach})
\begin{equation}\label{e.8}
   w_{-}(n)=\frac{1}{\tau (n)}:=\frac{1-\phi(n)}{\tau _{\infty }}
   + \frac{\phi(n)}{\tau _0}\,,
\end{equation}
where the value $\tau (n)$ can be interpreted as the characteristic time needed
for the first car in the cluster to leave it and to go out from its downstream
boundary at a distance about the headway distance $h_{\text{free}}(n)$ in the
current ``free'' flow state. The function $\phi(n)$ allows for the dependence
of the detachment time $\tau (n)$ on the cluster size $n$. We note that
expression~(\ref{e.8}) is the main original part of the model under
consideration.

When the cluster is sufficiently large, $n\gg 1$, it is reasonable to regard
the characteristic time $\tau(n)\approx\tau_{\infty}$ as a constant (i.e. $\phi
(n)\rightarrow 0$ for $n\gg 1$) as was done in papers~\cite{M1,M2} for all the
values of $n$.

For small clusters the $\tau (n)$-dependence, however, requires a special
attention. The matter is that the car attachment rate $w_{+}(n)$ is considered
to be directly determined by the local characteristics of the ``free'' flow
phase and the car cluster. Thus the dependence of the attachment rate
$w_{+}(n)$ on the cluster size $n$ arises via the headway distance
$h_{\text{free}}(n)$ in the ``free'' flow being a function of $n$, i.e.
$w_{+}(n)=w_{+}\left[h_{\text{free}}(n),h_{\text{clust}}\right]$. Therefore the
attachment rate is actually an explicit function $w_{+}(\rho ,\eta )$ of the
mean car density $\rho $ and the cluster relative volume $\eta$ only and, so,
exhibits minor variations on scales $\delta n\ll N$. As will be seen below,
exactly this feature is essential rather than the particular form of
$w_{+}(\rho,\eta )$ given here for definiteness sake only, because to describe
traffic breakdown at least one of the kinetic coefficients $w_{+}(n)$ and
$w_{-}(n)$ has to be a direct function of the cluster size $n$ for its
relatively small values corresponding to the formation of the cluster critical
nucleus. We associate this dependence with the escaping rate $w_{-}(n)$ that,
in contrast to the attachment rate $w_{+}(n)$, exhibits substantial variations
in the region $n\lesssim n_{0}\sim 20$.

The parameter $n_{0}$ actually divides the car clusters into the large cluster
group, $n\gg n_0$, for which the escaping rate is constant, $\phi(n)\rightarrow
0$, and the group of small clusters, $n\lesssim n_{0}$, whose dissolution is
affected substantially by the size $n$. This assumption is based on the fact
that there should a variety of possible manoeuvres for a driver to escape from
a sufficiently small cluster on a multilane highway when the lanes are not too
crowded.

Expression~(\ref{e.8}) takes into account this effect via the function
$\phi(n)$ running from 1 to 0 as the cluster size $n$ increases, so,
$\phi(1)\simeq 1$ and $\phi(n)\rightarrow 0$ as $n\rightarrow\infty$. In
particular, for a small neighborhood of the precluster size, $n\sim1$, the
value $\tau_0$ of $\tau(n)$ gives us actually the lifetime of the preclusters
and is assumed to be less than the escaping time from a large cluster, i.e.
$\tau_0<\tau_{\infty}$. Naturally, for the case of no cluster on the road we
have to set $w_{-}(0)= 0$. The main results will be obtained below actually
applying to the general properties of the dependence $w_{-}(n)$, however for
simplicity sake, we will adopt the following ansatz for $n\geq 1$
\begin{equation}\label{phi}
 \phi (n):= \left.\phi [x]\right|_{x = \tfrac{n}{n_0}}:= \frac{1}{(1+x)^q} \,,
\end{equation}
where the exponent $q>1$ is regarded as a given constant. We point out once
more that the dependence of the characteristic time $\tau (n)$ on the cluster
size is crucial because it is responsible for the existence of the metastable
``free'' flow phase.

The system of equations~(\ref{e.1})--(\ref{e.3}) subject to the initial
condition~(\ref{e.i}) with the relationships~(\ref{e.5}), (\ref{e.7}), and
(\ref{e.8}) forms the proposed probabilistic model for the car aggregation.
Within this model we will analyze the characteristic features of the large
cluster emergence and the form of the fundamental diagram, i.e. the ``flow
volume -- car density'' relation in the vicinity of traffic breakdown. In
particular, in the adopted terms the flow volume $j(n)$ for the given traffic
flow state, i.e. when a cluster of size $n$ arises on the road is written as
\cite{M1,M2}:
\begin{equation}
j(n)=(1-\eta )\rho \vartheta \left[ h_{\text{free}}(n)\right] +\eta \rho
\vartheta \left[ h_{\text{clust}}\right] \,.  \label{fd}
\end{equation}
Averaging expression~(\ref{fd}) with respect to the distribution
${\mathcal{P}}(n,t)$ we will get the fundamental diagram $j=j(\rho )$.

\subsection{Equilibrium distribution}

To clarify the characteristic features of the cluster formation let us analyze,
first, the stationary cluster sizes distribution $\mathcal{P}_{\text{eq}}(n)$.
The system of equations~(\ref{e.1})--(\ref{e.3}) subject to the initial
condition~(\ref{e.i}) admits the stationary solution
$\mathcal{P}_{\text{eq}}(n)$ meeting the zero ``probability'' flux in the
cluster size space:
\begin{equation*}\label{KuhAdv1}
    w_+(n-1)\mathcal{P}_{\text{eq}}(n-1) - w_-(n)\mathcal{P}_{\text{eq}}(n)
    = 0\,.
\end{equation*}
Whence we see that
\begin{equation*}\label{KuhAdv2}
  \frac{\mathcal{P}_{\text{eq}}(n)}{\mathcal{P}_{\text{eq}}(n-1)}
  = \frac{w_+(n-1)}{w_-(n)}\,,
\end{equation*}
enabling us to write the expression
$$
\mathcal{P}_{\text{eq}}(n)\propto\exp\left\{-\Omega (n)\right\}\,,
$$
where the function $\Omega (n)$ (called below the car growth potential) is
specified for $n\geq 2$ by the formula
\begin{eqnarray}
 \nonumber
 \Omega (n)& = & {}-\sum_{n^{\prime }=1}^{n-1}\ln \left[ \tau _{\infty
 }w_{+}^{\text{ov}}(n^{\prime })\right]\\
 &&{}+ \sum_{n^{\prime }=2}^{n}\ln \Bigl[
 1+\frac{(\tau_\infty-\tau_0)}{\tau_0}\,\phi (n^{\prime })\Bigr] \,.  \label{e.9}
\end{eqnarray}
Both of the terms in (\ref{e.9}) vary weakly as the argument $n$ changes by
one, enabling us to convert sum~(\ref{e.9}) into an integral with respect to
the cluster size $n$ treated as a continuous variable:
\begin{align}
\Omega (n) & = \Omega_{\infty} (n)+\Omega_0(n)\label{e.9a}\\
\intertext{where}%
\Omega_{\infty}(n) & \simeq -\int_{0}^{n}dn^{\prime }\, \ln \left[ \tau
_{\infty }w_{+}^{\text{ov}}\left[ h_{\text{free}}(n^{\prime })\right]
\right]\label{e.9a1}\,,\\
\Omega_{0}(n) & \simeq \hphantom{-} \int_{0}^{n}dn^{\prime }\,\ln \Bigl[
1+\frac{(\tau_\infty-\tau_0)}{\tau_0}\,\phi (n^{\prime })\Bigr] \,.
\label{e.9a2}
\end{align}

The former term in (\ref{e.9}) or (\ref{e.9a}), i.e. the component
$\Omega_{\infty}(n)$ called below the growth potential mainly characterizes
whether a stable car cluster can arise on the road under the given conditions
and specifies its size because it exhibits substantial variations on large
scales exceeding substantially the size $n_0$. By contrast the latter one, the
component $\Omega_{0}(n)$, describes the formation of the critical cluster
nucleus and, so, the breakdown phenomenon. Indeed, as follows from (\ref{phi})
and (\ref{e.9a2}) the potential $\Omega_{0}(n)$ is constant for $n\gg n_0$ and,
thus, cannot affect the growth of a large cluster already formed on the road.
Besides, within the continuum approximation we have ignored the details of the
cluster distribution in the region including both the points $n=0$ and $n=1$
and expand the cluster space $n\geq 1$ to the whole axis $n\geq 0$.

\begin{figure}
\begin{center}
\includegraphics[scale=1]{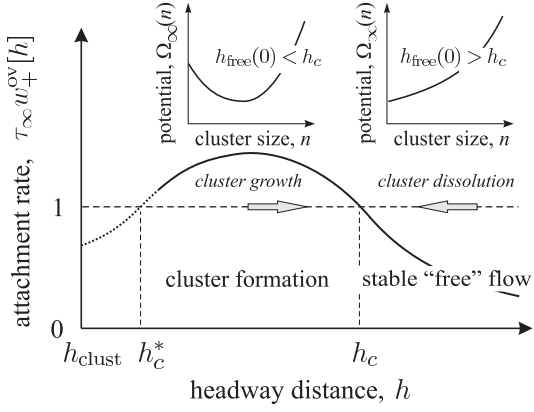}
\end{center}
\caption{The attachment rate $w_{+}^{\text{ov}}\left[ h\right]$ \textsl{vs} the
headway distance $h$ and the stability regions of the ``free'' flow
phase.\label{schem}}
\end{figure}

Let us, first, analyze the condition of the cluster emergence. Applying to
Fig.~\ref{schem} we can see that a large cluster can arise on the road, in
principle, if there exists a value of the headway distance $h_{c}$ meeting the
equality
\begin{equation}\label{Hcdef}
\tau _{\infty }w_{+}^{\text{ov}}\left[ h_{c}\right] =1\,,
\end{equation}
which will be assumed to hold beforehand. In particular, within
approximation~(\ref{e.5}) together with (\ref{e.4}) for $D_{\text{opt}}\gg
h_{\text{clust}}$ and $p=2$ this assumption holds if $\tau _{\infty
}v_{\text{max}}>2D_{\text{opt}}$ and the critical headway reads
\begin{equation*}
h_{c}^{(2)}=\frac{1}{2}\left( \tau _{\infty }v_{\text{max}}+\sqrt{(\tau
_{\infty }v_{\text{max}})^{2}-4D_{\text{opt}}^{2}}\right)\,,
\end{equation*}
whereas for $p \to \infty$ and $\tau _{\infty }v_{\text{max}}>D_{\text{opt}}$
we have
\begin{equation*}
h_{c}^{(\infty )}=\tau _{\infty }v_{\text{max}}\,.
\end{equation*}
The ``free'' flow phase will be stable if the initial headway distance
$h_{\text{free}}(0) > h_c$ and unstable otherwise.

Let us justify these statements. The growth potential $\Omega(n)$ is actually
the sum of $\ln\left[w_-(n)/w_+(n)\right]$ over $n$ (see formula~(\ref{e.9})).
So, in the region where the integrand of (\ref{e.9a1}) meets the inequality
$\tau _{\infty}w_{+}^{\text{ov}}\left[ h\right] <1$ and the potential
$\Omega_{\infty}(n)$ is an increasing function of $n$, the cluster dissolution
is more intensive then the car attachment. Under these conditions the cluster
size on the average decreases in time. The same concerns the time dependence of
the headway distance $h_{\text{free}}(n)$ in the ``free'' flow phase because
the value of $h_{\text{free}}(n)$ decreases as the cluster becomes smaller
(Fig.~ \ref{dvss}), which is also illustrated by arrows in Fig.~\ref{schem}.
Since $\tau _{\infty} w_{+}^{\text{ov}}\left[ h\right] <1$ for $h>h_c$ any
randomly arising cluster tends to disappear and, consequently, the ``free''
flow phase is stable when $h_{\text{free}}(0) > h_c$. In this case the
potential $\Omega_{\infty}(n)$ possesses one minimum located at the boundary
point $n=0$ (or $n=1$ what is the same in the continuum description).

Otherwise, $h_{\text{free}}(0) < h_c$, there is a region
$h_{\text{free}}(0)<h<h_c$ where $\tau _{\infty}w_{+}^{\text{ov}}\left[
h\right] >1$ and the car attachment rate exceeds that of the cluster
dissolution and a cluster occurring in the corresponding ``free'' flow state
tends to grow, inducing the further increase in the headway distance
$h_{\text{free}}(n)$. In this case the ``free'' flow phase is unstable and the
cluster will continue to grow until the value of $h_{\text{free}}(n)$ reaches
the critical point $h_c$, where the car attachment and the cluster dissolution
balance each other. Whence it follows, in particular, that the developed
cluster is of the size $n_{\text{clust}}$ obeying the equation
\begin{equation}
h_{\text{free}}(n_{\text{clust}})=h_c \label{z1}
\end{equation}
and the $\Omega_{\infty}(n)$ has a minimum at the internal point
$n=n_{\text{clust}}$. In the present paper we will ignore the existence of
another region where the equality $\tau _{\infty }w_{+}^{\text{ov}}\left[
h\right] <1$ also holds for very dense traffic flow, which has been considered
in papers~\cite{M1,M2}.

Relationship~(\ref{e.6}) enables us to rewrite the instability conditions in
terms of the mean car density $\rho$. The critical value $\rho_c$ of the car
density is the solution of the equation $h_{\text{free}}(0)=h_c$, whence we
immediately get
\begin{equation}\label{e.10}
\rho_{c1}=\rho _{\text{lim}}\frac{l_{\text{car}}+h_{\text{clust}}}
{l_{\text{car}}+h_{c}}\,.
\end{equation}
Then the stable state of the ``free'' flow corresponds to the inequality
$\rho<\rho_{c1}$ and it loses the stability when $\rho>\rho_{c1}$. In the
latter case a large cluster arises on the road whose size
$n_{\text{clust}}(\rho )=\eta_{\text{clust}}(\rho )N$ and relative volume
\begin{equation}
\eta_{\text{clust}}(\rho
)=\frac{h_{c}+l_{\text{car}}}{h_{c}-h_{\text{clust}}}\cdot \frac{\rho -\rho
_{c1}}{\rho }\,.  \label{e.11}
\end{equation}

\begin{figure}
\begin{center}
\includegraphics{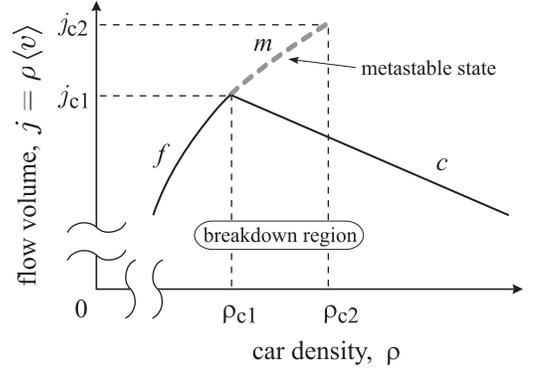}
\end{center}
\caption{The fundamental diagram in the vicinity of the breakdown
region.\label{Ffd}}
\end{figure}

In the given analysis we have ignored the dependence of the cluster
dissolution rate $w_-(n)$ on the size $n$ and, thereby, the considered picture
describes actually the ``free'' flow--cluster transition of the second order.
It does not allow for the metastable state of the ``free'' flow phase and
corresponds to the continuous transition from the traffic state of no cluster
on the road to the formation of a certain cluster whose relative volume
changes continuously from zero as the car density penetrates deeper in the
instability region (see formula~(\ref{e.11})). Consequently, this
approximation cannot explain  the traffic breakdown and on the phase diagram
matches solely the stable branches \textit{``f''} and \textit{``c''} of the
``free'' flow and the traffic with a developed cluster, respectively
(Fig.~\ref{Ffd}).
Nevertheless, exactly the given approximation describes the stable branches of
the fundamental diagram and, moreover, the metastable branch is a continuation
of the branch \textit{``f''} into the instability region. Keeping the latter
in mind we present also the expression specifying these branches:
\begin{equation*}
\label{branch} j_{\text{\it fc}}(\rho )=\left\{
\begin{array}{lll}
\rho \vartheta \left[ h_{c}+(h_{c}+l_{\text{car}})(\rho _{c1}-\rho )/\rho
\right]  &\ \text{if} & \rho <\rho _{c1}\,, \\
j_{c1}- G\left(\rho -\rho_{c1}\right)/\rho_{c1} &\ \text{if}& \rho
>\rho _{c1}\,,
\end{array}
\right.
\end{equation*}
where the constants
\begin{eqnarray*}
j_{c1}&=&\rho_{c1}\vartheta \left[ h_{c}\right], \\
G&=&\frac{\left(h_{c}+l_{\text{car}}\right) }{\left(
h_{c}-h_{\text{clust}}\right)} \frac{\rho_{c1}}{\rho_{\text{lim}}}\left(
\rho_{c1}\vartheta \left[ h_{c}\right] -\rho_{\text{lim}}\vartheta \left[
h_{\text{clust}}\right] \right).
\end{eqnarray*}
It should be noted that in obtaining this expression we have substituted the
maximum probability value $n_{\text{clust}}$ of the cluster size into
expression~(\ref{fd}) instead of averaging it over the distribution
$\mathcal{P}_{\text{eq}}(n)$. The latter is justified because the effect of
the cluster size fluctuations is ignorable due to $N\gg 1$.

Now we analyze possible metastable states of the ``free'' flow phase. In order
to do this we should take into account both the component of the growth
potential $\Omega(n)$. Since the function $\Omega_{0}(n)$ exhibits remarkable
variations in the region $n\lesssim n_0$ only and, thus, the size $n_c$ of the
critical nucleus also belongs to this region we may confine our consideration
to clusters whose size $n$ is much less than the final cluster size
$n_{\text{clust}}$ attained after the instability development. In addition for
the sake of simplicity we will regard the value
$(\tau_\infty-\tau_0)/\tau_\infty$ as a small parameter, which enables us to
examine solely a small neighborhood of the instability boundary,
$0<\rho-\rho_{c1} \ll \rho_{c1}$.

In this case the value of $\tau _{\infty }w_{+}^{\text{ov}}\left[
h_{\text{free}}(n)\right]$ is practically constant and can be approximated by
the expression
\begin{equation}
\ln\left\{\tau _{\infty
}w_{+}^{\text{ov}}\left[h_{\text{free}}(n)\right]\right\} \simeq  g\frac{\rho
-\rho _{c1}}{\rho _{c1}}\,,  \label{e.12}
\end{equation}
where the coefficient
\begin{equation*}
g=\frac{\left( l_{\text{car}}+h_{c}\right) }{h_{c}}\left| \frac{d\ln w_{+}^{
\text{ov}}[h]}{d\ln h}\right| _{h=h_{c}}
\end{equation*}
is about unity, $g\sim 1$, in the general case. In particular, for the stepwise
dependence $\vartheta (h)$ (if we set $p=\infty $ in expression~(\ref{e.4}))
and $D_{\text{opt}}\gg l_{\text{car}}$, $h_{\text{clust}}$ we have the rigorous
equality $g=1$. Expression~(\ref{e.12}) together with formula~(\ref{phi})
allows us to represent the dependence of the growth potential $\Omega(n)$ on
the cluster size $n$ as
\begin{eqnarray}
\frac{d\Omega(n)}{dn}&\approx &
  \frac{(\tau_{\infty }-\tau _{0})}{\tau_0}\,\phi(n)-
  g\frac{(\rho -\rho _{c1})}{ \rho _{c1}}\nonumber\\
 & =& \frac{(\tau_{\infty }-\tau _{0})}{\tau_0} \left(\frac{n_0}{n_0+n}\right)^q
- g\frac{(\rho -\rho _{c1})}{ \rho _{c1}}\,.\quad \label{z2}
\end{eqnarray}
The first term on the right-hand side of (\ref{z2}) is due to the increase in
the cluster dissolution rate for $n\lesssim n_0$, whereas the latter one is
proportional to the cluster growth rate in the region of large values of $n$.
The resulting value of the derivative $d\Omega(n)/dn$ characterizes the
direction of the cluster evolution. If it is positive, $d\Omega(n)/dn> 0$, i.e.
the potential $\Omega(n)$ is an increasing function of $n$ the cluster
dissolution is the dominant process and the cluster of size $n$ tends to
dissipate. Otherwise, i.e. when $d\Omega(n)/dn< 0$ it will grow.

\begin{figure}
\begin{center}
\includegraphics[scale=1]{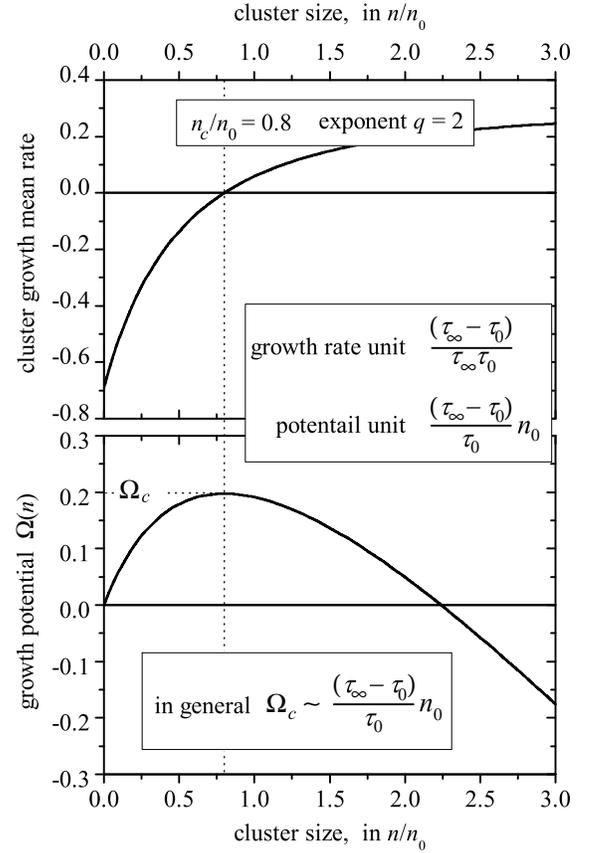}
\end{center}
\caption{The form of the cluster growth potential $\Omega(n)$ in the breakdown
region, (\textit{upper window}) the value of $-d\Omega(n)/dn$ proportional to
the mean rate of cluster growth \textsl{vs} the cluster size $n$ and
(\textit{lower window}) the growth potential $\Omega(n)$ itself. The present
figures have been obtained using ansatz~(\protect\ref{phi}) with the exponent
$q = 2$ and the chosen value of the vehicle density $\rho$ gives the ratio of
the critical nucleus size $n_c$ to the characteristic value $n_0$ equal to
$n_c/n_0 = 0.8.$ \label{2in1}}
\end{figure}

The former term attains its maximum at $n=0$, so, according to (\ref{z2}) the
derivative $d\Omega(n)/dn$ is negative for all the possible values of the
cluster size under consideration $0\leq n\ll n_{\text{clust}}$ when
\begin{equation}
\rho > \rho_{c2}:=\rho _{c1}\left[ \frac{\left( \tau _{\infty }-\tau
_{0}\right) }{g\tau _0}+1\right]\,.   \label{e.14}
\end{equation}
In this case the ``free'' flow phase becomes absolutely unstable. Under the
opposite condition, $\rho_{c1}<\rho<\rho_{c2}$ there is a certain value $n_c$
at which the derivative $d\Omega(n)/dn$ changes the sign (Fig.~\ref{2in1}).
Setting the left-hand side of (\ref{z2}) equal to zero we get the relationship
\begin{align}
 \label{z3a}
 \phi(n_c) & = \frac{\rho-\rho_{c1}}{\rho_{c2}-\rho_{c1}}\\
 \intertext{which togather with ansatz (\protect\ref{phi}) gives the estimate}
 \label{z3b}
 n_{c} & = n_{0}\left[\left(\frac{\rho_{c2}-\rho _{c1}}{\rho-\rho _{c1}} \right)
 ^{\tfrac1q}  - 1 \right]\,,
\end{align}
well justified except for small neighborhoods of the boundary points
$\rho_{c1}$ and $\rho _{c2}$. If $n<n_c$ the derivative is positive and the
cluster should decrease in size, i.e. the ``free'' flow phase is stable with
respect to the emergence of such small clusters. However, if a certain cluster
of size $n>n_c$ has already formed, for example, due to random fluctuations,
then it will grow and a large cluster of size $n_{\text{clust}}$ arises on the
road because $d\Omega(n)/dn<0$ in the region $n>n_c$.

In other words, we have shown that the dependence of the dissolution rate
$w_-(n)$ on the cluster size $n\lesssim n_0$ makes the ``free'' flow phase
metastable when the car density belongs to the interval
$\rho\in(\rho_{c1},\rho_{c2})$ (branch \textit{``m''} in Fig.~\ref{Ffd}). The
formation of a large cluster, $n\gg n_0$, proceeds via generation of the
critical nucleus whose size $n_c$ is estimated by expression~(\ref{z3b}). In
order to find the generation rate of the critical nuclei and, thus, the
breakdown frequency we should consider the transient processes in the cluster
growth, which is the subject of the next section.

\subsection{Continuum approximation. The breakdown probability\label{sec:FPE}}

In order to apply well developed techniques of the escaping theory (see, e.g.,
\cite{Gard}) to the analysis of the traffic breakdown probability we
approximate the discrete master equation~(\ref{e.1}) by the corresponding
Fokker-Planck equation. It is feasible because in the case under consideration
the kinetic coefficients $w_{+}(n)$, $w_{-}(n)$, first, vary smoothly on scales
about unity and, second, are approximately equal to each other, $\left|
w_{+}(n)-w_{-}(n)\right| \ll w_{-}(n)$. The latter conditions enable us to
treat the argument $n$ as a continuous variable and to expand the functions
$w_{+}(n\pm 1)$, $w_{-}(n\pm 1)$, and $\mathcal{P}(n\pm 1,t)$ into the Taylors
series. In this way and, in addition, taking into account
expression~(\ref{e.8}) we reduce equation~(\ref{e.1}) to the following
Fokker-Planck equation
\begin{equation}\label{e.FP}
    \tau _{\infty }\partial_t \mathcal{P}(n,t) = \partial_n
    \left[\partial_n \mathcal{P}(n,t)+
    \mathcal{P}(n,t)\partial_n\Omega (n)\right] ,
\end{equation}
where the potential $\Omega (n)$ is given by formula~(\ref{e.9a}) in the
general form. However, in the case under consideration the ratios $n/N$,
$(\rho-\rho_{c1})/\rho_{c1}$, and $(\tau_\infty -\tau_0)/\tau_0$ are regarded
to be sufficiently small and it is possible to expand the potential $\Omega(n)$
in the tree parameters and to remain the leading terms only. In this way we get
\begin{multline}\label{Om}
   \Omega (n) = \frac{(\tau_{\infty}-\tau_0)}{\tau_0} n_{0}\\
   {}\times\Bigg\{
   \int_0^{\tfrac{n}{n_0}} dx\,\phi[x]
   - \phi[x_c]\frac{n}{n_0}\Bigg[1-\frac{n}{2n_{\text{clust}}(\rho)}\Bigg]
   \Bigg\}\,,\
\end{multline}
where $x_c = n_c/n_0$ and we have set $\Omega(0)=0$. Equation~(\ref{e.2})
transforms into the boundary condition at infinitely distant points that is
imposed on the probability flux
$$
    J(n):=-\partial_n\mathcal{P}(n,t) -
    \mathcal{P}(n,t)\partial_n\Omega(n)
$$
and requires it to be equal to zero, $J(\infty)=0$. Equation~(\ref{e.3})
describing the precluster generation is reduced, in turn, to the zero boundary
condition imposed on the probability flux $J(n)$ formally at $n=0$, i.e.
$J(0)=0$. The latter is justified by the assumed quasi-equilibrium between the
``free'' flow phase and the preclusters. And, finally,   the initial
condition~(\ref{e.i}) can be rewritten as
$$
    \int_0^\infty dn\,\mathcal{P}(n,t) =1.
$$

When the car density belongs to the interval $\rho\in(\rho_{c1},\rho_{c2})$ and
the ``free'' flow phase is metastable the form of the growth potential $\Omega
(n)$ in the region $n\lesssim n_0$ is shown in Fig.~\ref{2in1}. The the
``free'' flow phase being in quasi-equilibrium with preclusters matches the
local minimum at $n=0$ separated from the region of the stable cluster growth
$n> n_c$ by the potential barrier $\Omega_{c}$. The value of this the potential
barrier is estimated as
\begin{align}\label{asa1}
  \Omega _{c}& \simeq \frac{(\tau_{\infty}-\tau_0)}{\tau_0} n_{0}\,
  \omega\Big[\text{\large$\tfrac{n_c}{n_0}$}\Big]\\
  \intertext{where the function}
  \omega[x_c] & :=
  \int_0^{\tfrac{n_c}{n_0}} dx\,x\left(-\frac{d\phi[x]}{dx}\right)\,.
  \label{nonum}
\end{align}
In particular, for ansatz~(\ref{phi}) with the exponent $q=2$ it
expression~(\ref{asa1}) becomes
\begin{equation}\label{asa11}
  \Omega _{c}\simeq \frac{(\tau_{\infty}-\tau_0)}{\tau_0} n_{0}
  \frac{x_c^2}{(1+x_c)^2}\,,
\end{equation}
moreover, in the limit $x_c\ll 1$ we have
$$
  \omega[x_c]\simeq \frac{1}{2}r x^2_c\,,\ \text{where the constant}\
  r = -\left.\frac{d\phi[x]}{dx}\right|_{x=0}\;,
$$
as follows from expression~(\ref{nonum}), and the general formula for the
potential $\Omega_c$ can be written as
\begin{align}\label{yyy1}
  \Omega_{c} & \simeq \frac{(\tau_{\infty}-\tau_0)}{2\tau_0}\,r n_{0}
  x_c^2\,.\\
  \intertext{In the same limit expression~(\ref{z3a}) gives us}
  x_c & \simeq \frac{(\rho_{c2}-\rho)}{r(\rho_{c2}-\rho_{c1})}\,.
  \label{yyy11}
\end{align}
The main much deeper minimum of the potential $\Omega (n)$ is located at
$n=n_{\text{clust}}\gg n_0$.

We have demonstrated that a precluster must climb over the potential barrier
$\Omega _{c}$ at the point $n_{c}$ to convert into a large stable cluster. It
is implemented through random fluctuations carrying the cluster size up to the
critical value $n_{c}$. In this terms the traffic breakdown is the classical
escaping from a potential well described by the Fokker-Planck
equation~(\ref{e.FP}). The latter analogy enables us to write down the estimate
for the frequency $\nu_{\text{bd}}$ of the traffic breakdown processes
depending on the given vehicle density in the ``free'' flow state. Namely, as
shown in Appendix
\begin{multline}\label{e.final}
    \nu _{\text{bd}}\simeq \frac{1}{\sqrt{2\pi n_0}\tau_{\infty}}
    \Big(\frac{\tau_{\infty}-\tau _{0}}{\tau_0}\Big)^{\tfrac32}
    (1-\phi[x_c])\left|\phi'[x_c]\right|^{\tfrac12}\\
    {}\times\exp\left\{-\frac{(\tau_{\infty}-\tau_0)}{\tau_0}n_{0}\,\omega[x_c]\right\}\,,
\end{multline}
which is well justified for the car density $\rho$ belonging to the interval
$\rho _{c1}<\rho <\rho _{c2}$ except for a certain sufficiently small
neighborhoods on the critical points $\rho _{c1}$, $\rho _{c2}$.
Ansatz~(\ref{phi}) with the exponent $q=2$ together with formula~(\ref{z3a})
enables us to rewrite expression~(\ref{e.final}) as
\begin{multline}\label{e.final1}
    \nu _{\text{bd}}\simeq \frac{1}{\sqrt{\pi n_0}\tau_{\infty}}
    \Big(\frac{\tau_{\infty}-\tau _{0}}{\tau_0}\Big)^{\tfrac32}
    (1-\Delta)\Delta^{\tfrac34}\\
    {}\times\exp\left\{-\frac{(\tau_{\infty}-\tau_0)}{\tau_0}n_{0}
    \frac{(1-\Delta)^2}{\big(1+\Delta^{\frac12}\big)^{2}}
    \right\}\,.
\end{multline}
Here we have introduced the quantity
\begin{equation}\label{e.delta}
    \Delta := \frac{\rho-\rho_{c1}}{\rho_{c2}-\rho_{c1}}
\end{equation}
treated as a dimensionless overcriticality measure showing how deep the system
penetrates into the metastability region, $\Delta = 0$ corresponds to the value
$\rho_{c1}$ of the vehicle density where a jam can emerge in principle and
$\Delta = 1$ matches the vehicle density $\rho_{c2}$ after exceeding which no
traffic states except for jams can exist at all (Fig.~\ref{Ffd}).

\subsection{Frequency of traffic breakdown during a fixed time interval}

Experimentally traffic breakdown is typically analyzed detecting a significant
drop in the vehicle speed during a certain fixed time interval $T_{\text{obs}}$
about several minutes and then drawing the relative frequency of these events
\textsl{vs} the traffic volume \cite{Br1,Br2,Br3,Br4}. In order to compare this
representation with the obtained results let us consider them in more details.

As follows from expression~(\ref{e.14}) the density interval
$(\rho_{c1},\rho_{c2})$ inside which the traffic jam emerges by the nucleation
mechanism is of the thickness
$$
(\rho_{c2}-\rho_{c1})=\rho_{c1}\frac{(\tau_\infty-\tau_0)}{g\tau_0}\,.
$$
According to the experimental data \cite{Br1,Br2,Br3,Br4,KBr} the thickness of
the traffic volume interval inside which the traffic breakdown demonstrates the
probabilistic behavior is about its low boundary in magnitude. So we have to
regard the ratio $(\tau_\infty-\tau_0)/\tau_0$ also as a value about unity:
\begin{equation}\label{estim.1}
    \frac{(\tau_\infty-\tau_0)}{g\tau_0}\sim 1\,.
\end{equation}
Thereby, setting $n_0=20$ we get the conclusion that in the general case where
$n_c\sim n_0$ the potential barrier $\Omega_c\sim 5$ corresponding to the
exponential factor $\exp\{-\Omega_c \}\sim 0.7\times 10^{-2}$. Then setting
$\tau_\infty\sim 2$~sec and estimating the preceding cofactor as $1/(\sqrt{2\pi
n_0}\tau_{\infty})$ we find the characteristic rate of the traffic breakdown
being about 1/50 min${}^{-1}$ in the general case. So the real traffic
breakdown events seem to be observed in cases where the vehicle density comes
to the upper boundary $\rho_{c2}$. The latter allows us to confine our analysis
formally to the limit case
\begin{equation}\label{limit}
  x_c\ll 1 \Leftrightarrow  (\rho_{c2}-\rho)\ll (\rho_{c2}-\rho_{c1})\,.
\end{equation}

Then estimating the probability $\mathcal{F}_{\text{bd}}$ of detecting a
traffic breakdown during the observation time interval $T_{\text{obs}}$ as
$\mathcal{F}_{\text{bd}}= T_{\text{obs}}\nu_{\text{bd}}$ we obtain from
(\ref{e.final}) the expression
\begin{gather}\label{007}
    \mathcal{F}_{\text{bd}}(\Delta) = \frac{T_{\text{obs}}}{\tau_{\text{bd}}}\,
    2\,\Omega_c^{\frac12}\exp\left\{- \Omega_c \right\}\,,\\
    \intertext{where}
    \Omega_c = \frac{(\tau_\infty -\tau_0)}{2r\tau_0}n_0(1-\Delta)^2
    \intertext{and we have introduced the time scale}
    \tau_{\text{bd}} =
    \frac{2\sqrt{\pi}\tau_0}{r(\tau_\infty-\tau_0)}n_0\tau_\infty\,,
    \label{bdtime}
\end{gather}
giving us the characteristic time of the breakdown emergence. In deriving
(\ref{007}) we have also taken into account formulae~(\ref{phi}), (\ref{yyy1}),
and (\ref{yyy11}) and remained directly number 2 as cofactor because the
maximum of the function $z^{1/2}\exp(-z)$ is about 0.43. Naturally, we have to
confine ourselves to such values of the vehicle density for which
$\mathcal{F}_{\text{bd}}\leq 1$ because traffic flow with higher values of the
vehicle density cannot exist on these time scales. For the following values
$r=2$, the ratio $(\tau_\infty-\tau_0)/\tau_0\sim 1$, $\tau_\infty\sim 2$~sec,
and $n_0\sim 20$ expression~(\ref{bdtime}) gives us the estimate
$\tau_{\text{bd}}\sim 1$~min of the characteristic breakdown time. It should be
pointed out that the latter estimate does not contradict the evaluation of the
breakdown rate given at the beginning of the present section because it holds
only in the region $\Omega_c\gg1$.

\begin{figure}[!]
\begin{center}
\hspace*{10mm}ansatz (\ref{phi}) with the exponent $q = 2$\\
\includegraphics{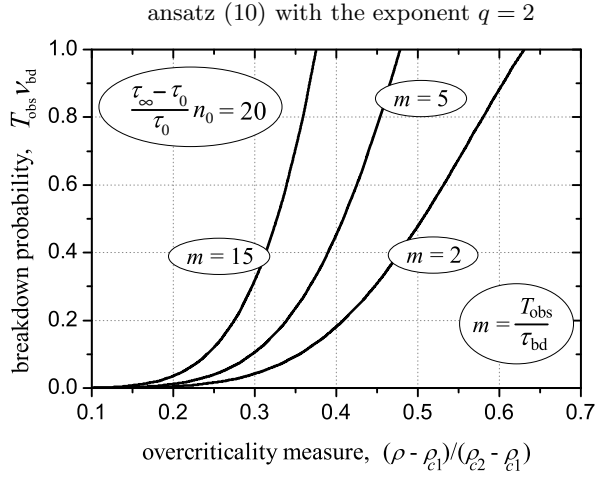}
\end{center}
\caption{The traffic breakdown probability \textsl{vs} the depth
$(\rho-\rho_{c1})/ (\rho_{c2}-\rho_{c1})$ of penetration into the metastability
region.\label{Fbd}}
\end{figure}

Figure~\ref{Fbd} illustrates the obtained results depicting the breakdown
probability for different values of the observation time $T_{\text{obs}}$
measured in units of $\tau_{\text{bd}}$ \textsl{vs} the depth of penetrating
into the metastability region. It should be pointed out that in drawing
Fig.~\ref{Fbd} we have applied to formula~(\ref{e.final1}) rather than
(\ref{007}) in order to have a possibility to go out of the frameworks of the
formal limit~(\ref{limit}). The latter is considered hear to clarify the
obtained results only. To make the form of the
$\mathcal{F}_{\text{bd}}(\Delta)$-dependence more evident we apply again to the
formal limit case~(\ref{limit}) assuming the ratio
$m:=T_{\text{obs}}/\tau_{\text{bd}}$ to be a large parameter. Than analyzing a
small neighborhood of the point $\rho^*_m$ specified by the equality
\begin{gather}\label{t1}
  1 - \Delta^*_m \approx \left[\frac{2r\ln(2m)\tau_0}{(\tau_\infty-\tau_0)n_0}
  \right]^{\tfrac{1}{2}}\\
  \intertext{we get}
  \label{t2}
  \mathcal{F}_{\text{bd}}(\rho) =
  \exp\left(\frac{\rho-\rho^*_m}{\overline{\rho}_m}\right)\,,\\
  \intertext{where the vehicle density scale}
  \label{t3}
  \overline{\rho}_m = (\rho_{c2}-\rho_{c1})\left[\frac{r\tau_0}{2\ln(2m)(\tau_\infty-\tau_0)n_0}
  \right]^{\tfrac{1}{2}}\,.
\end{gather}
Therefore, in a rough approximation the
$\mathcal{F}_{\text{bd}}(\rho)$-dependence is a simple exponential function
whose scale $\overline{\rho}_m$ is approximately a constant value (because the
function $ln(m)$ shows week variations for $m\gg1$). Changing the observation
duration $T_{\text{obs}}$ practically shift the cut-off point $\rho^*_m$ only.

\section{Comparison with experimental results}

Experimental investigations of the traffic breakdown regarded as a
probabilistic phenomenon have been carried out by several authors (see, e.g.
Ref.~\onlinecite{Br1,Br2,Br3,Br4,KuBr}). Elefteriadou \textit{et al.}
\cite{Br1} actually pointed out the fact that traffic breakdown at ramp merge
junctions occurs randomly without precise relation to a certain fixed value of
traffic volume. A more detailed analysis of the breakdown probability has been
fulfilled in papers~\cite{Br2,Br3,Br4,KuBr}. These observations show that
traffic breakdown can occur inside a wide interval of traffic volume from about
1500~veh/h/l (vehicles per hour per lane) up to 3000~veh/h/l. The real dynamics
of traffic breakdown near bottlenecks and the developed structure of the
congested traffic flow are sufficiently complex as it was exhibited by
Kerner~\cite{KBr}. In particular, Kerner demonstrated that the synchronized
mode of traffic flow in the vicinity of highway bottleneck is locally
metastable under the discharged downstream traffic flow of volume $j$ varying
in the same interval. The latter enables us to estimate the detachment time
$\tau_{\infty}$ playing the significant role in the presented model. In fact
ignoring the velocity $\vartheta(h_{\text{clust}})$ of cars in the cluster as
well as the headway distance $h_{\text{clust}}$ in ansatz~(\ref{e.5}) we get
that the lower boundary $\rho_{c1}$ of the metastability region meets the
traffic volume
$$
    j_{c1} = \rho_{c1}\vartheta(h_c)\approx w_{+}^{\text{ov}}[h_c] =
    \frac1{\tau_{\infty}} \sim 1800\ \text{veh/h/l}.
$$
Whence we find immediately the estimate $\tau_\infty \sim 2$~sec, which is in
agreement with the value adopted previously in papers~\cite{M1,M2}. In this
section, as it has been done in the previous one, we use the estimates of the
quantities $n_0\sim 20$ according to the experimental data depicted in
Fig.\ref{SVD}, set $(\tau_\infty-\tau_0)/\tau_0\sim 1$ from the general
consideration.

\begin{figure}
\begin{center}
\includegraphics{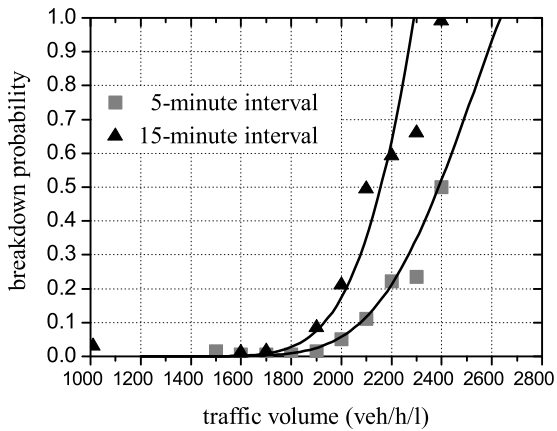}
\end{center}
\caption{The traffic breakdown probability during 5-minute and 15-minute
intervals \textsl{vs} the traffic volume (after Lorenz \&
Elefteriadou~\protect\cite{Br4}). The continuous curves present the
dependence~(\protect\ref{e.final1}) fitting these experimental data in the
frameworks of the replacement $\Delta \Leftarrow (j-j_{c1})/(j_{c2}-j_{c1})$
(see expression~(\protect\ref{e.delta})) under the following values of the
critical traffic volumes $j_{c1} = 1200$~veh/h/l and $j_{c2} = 3400$~veh/h/l,
the characteristic breakdown time $\tau_{\text{bd}}=2.5$~min, and the product
$n_0(\tau_\infty-\tau_0)/\tau_0 = 25$.\label{F:exper}}
\end{figure}

In order to compare the obtained results and the available experimental data we
have applied to the latest materials presented in detail by Lorenz \&
Elefteriadou \cite{Br4}. The breakdown phenomenon was investigated in traffic
flow near two bottlenecks of Highway~401, one of the primary Toronto traffic
arteries. The detectors were located right after the on-ramps within several
hundred meters downstream. So the dynamics of traffic breakdown observed at
these places seems to be mainly due to local internal properties of traffic
flow discussed in the present paper. The complex spatial structure of the
induced congested phase including moving wide and narrow jams reported by
Kerner~\cite{KBr} should emerge above the detectors upstream. We consider in
detail the data obtained for one these bottleneck (site ``A'' in
paper~\cite{Br4}). The paired detectors were located in each of the tree lanes
and were instrumented to provide vehicle count and speed estimates continuously
at 20-second intervals. A breakdown event was fixed via the velocity drop below
90~km/h, the middle point of a certain gap in the velocity field visually
separating the congested and free traffic flow states. Besides, only those
disturbances that caused the average speed over all the lanes to drop below
90~km/h for a period of five minutes or more were considered a true breakdown.
The latter enabled the authors to filter out large amplitude fluctuations in
the mean vehicle velocity not leading to the traffic breakdown.
Fig.~\ref{F:exper} exhibits the obtained probability (relative frequency) of
the traffic breakdown events during 5-minute and 15-minute intervals
\textsl{vs} the traffic volume partitioned within 100~veh/h/l steps. We note
that Fig.~\ref{F:exper} does not show the available 1~minute interval data
because the corresponding breakdown probability is not significant for all the
observed values of traffic volume except for the upper boundary 2800~veh/h/l,
that can be due to its rare occurrence.

The continuous curves in Fig.~\ref{F:exper} present our attempts to fit the
obtained theoretical dependence in the given experimental data. Namely these
curves describe the breakdown probability estimated as
${\mathcal{F}}_{\text{bd}} = T_{\text{obs}}\nu_{\text{bd}}$, where the latter
cofactor is given by the expression~(\ref{e.final1}) within the replacement
$\Delta \Leftarrow (j-j_{c1})/(j_{c2}-j_{c1})$ (see expression~(\ref{e.delta}))
and we have used the following values of the critical traffic volumes $j_{c1} =
1200$~veh/h/l and $j_{c2} = 3400$~veh/h/l, the characteristic breakdown time
$\tau_{\text{bd}}=2.5$~min (see expression~\ref{bdtime}), and set the product
$n_0(\tau_\infty-\tau_0)/\tau_0 = 25$. Keeping in mind the aforesaid all the
adopted values are quite reasonable including the estimate $j_{c2} =
3400$~veh/h/l seeming at first glance extremely high. Indeed, this value is no
more than a result of approximating the $j(\rho)$-dependence by a linear
function formally into the region of high vehicle densities.

\section{Summary and conclusion}

We have considered the traffic breakdown phenomenon regarded as a random
process developing via the nucleation mechanism. The origin of critical jam
nuclei proceeds in a metastable phase of traffic flow and seems to be located
inside a not too large region on a highway, for example, in the close proximity
of a highway bottleneck \cite{K5,K2}. The induced complex structure of the
congested traffic phase is located upstream the bottleneck \cite{KBr}. Keeping
these properties in mind we have applied to the probabilistic model regarding
the jam emergence as the development of a large car cluster on highway. In
these terms the traffic breakdown proceeds through the formation of a certain
car of critical size in the metastable vehicle flow, which enabled us to
confine ourselves to the single cluster model.

We assumed that, first, the growth of the car cluster is governed by attachment
of cars to the cluster whose rate is mainly determined by the mean headway
distance between the cars in the vehicle flow and, may be, also by the headway
distance in the cluster. Second, the cluster dissolution is determined by the
car escape from the cluster whose rate depends on the cluster size directly. To
justify the latter assumption we apply to the modern notion of the traffic flow
structure (see Ref.~\onlinecite{K1,K1a,K2}). Namely, the jam emergence goes
mainly through the sequence of two phase transitions: free flow $\rightarrow $
synchronized mode $\rightarrow $ stop-and-go pattern \cite{K6}. Both of these
transitions are of the first order, i.e. they exhibit breakdown, hysteresis,
and nucleation effects \cite{K5}. Therefore considering the final stage of the
jam emergence we have to regard the synchronized mode as the metastable phase
exactly inside which a critical jam nucleus appears due to random fluctuations.
The synchronized mode is characterized by strong multilane correlations in the
car motion and, as a result, all the vehicles in a certain affective cluster
spanning over all the highway lanes move as a whole. So the proposed
probabilistic description deals with actually macrovehicles comprising many
individual cars. The available single-vehicle experimental data \cite{N1}
present the correlation characteristics of the synchronized mode which have
enabled us to estimate the characteristic dimension $n_0\sim 20$--30 of the car
cluster entering the dependence of the car detachment rate on the cluster size.
Namely for small car clusters, $n\lesssim n_0$, the characteristic detachment
time $\tau_0$ should be substantially less than this time $\tau_\infty$ for
large clusters, $n\gg n_0$.

We have written the appropriate master equation for the cluster distribution
function and analyze the formation of the critical car cluster due to the climb
over a certain potential barrier. The inequality $n_0\gg 1$ has opened us the
way to convert from the discrete master equation to the appropriate
Fokker-Plank equation and find all the required characteristics of the traffic
breakdown.

The obtained results are compared with the available experimental data and, in
detail, with the probability of traffic breakdown in the vicinity of
bottlenecks \textsl{vs} the traffic volume presented by Lorenz \& Elefteriadou
\cite{Br4}. It turned out that the theoretical curves can be fitted closely to
the given experimental data using values of the main parameters chosen based on
the general properties of the traffic flow not related directly to the
breakdown dynamics. In particular, first, we have demonstrated that the
characteristic internal time scale $\tau_{\text{bd}}$ of the breakdown
development is about $\tau_{\text{bd}}\sim n_0\tau_\infty$ (we recall that
$\tau_\infty\sim 2$~sec is the characteristic time during which a car can
individually leave a cluster). Whence we get the estimate of the breakdown
timescale about one minute. The latter justifies the widely used probabilistic
technique of the breakdown investigation based on fixing this event during a
time interval of several minutes. Second, the proposed model explains why the
traffic breakdown as a probabilistic phenomenon is observed inside a
sufficiently wide interval of the traffic volume, namely, the thickness
$\triangle j$ of this layer can attain its low boundary $j_{c1}$ in magnitude.
The matter is that $\triangle j/j_{c1}\sim (\tau_\infty-\tau_0)/\tau_0\sim 1$.

Concluding the aforesaid we state that traffic breakdown is a mesoscopic
process, as it must be for the synchronized mode, whose characteristic spatial
and temporal scales correspond to car clusters made of a large number of
vehicles.

\begin{acknowledgments}
The authors would like to thank Peter Wagner (Berlin) for useful comments and
criticism.
\end{acknowledgments}

\appendix
\section{Escaping rate from a boundary well}

In section~\ref{sec:FPE} we have obtained the Fokker-Plank
equation~(\ref{e.FP}) governing the evolution of car clusters treated as random
wandering in the space of their size $n$. It has turned out that near the
threshold the precluster domain is separated from the large cluster region by a
potential barrier $\Omega_c$, so, the formation of the clustered phase should
proceeds through the nucleation mechanism (Fig.~\ref{2in1}). In other words,
for a large cluster to emerge on the road its critical nucleus $n_c$ has to
arise via random fluctuations of the cluster size in the precluster region.
Thereby in order to describe the cluster formation we need the expression
specifying the rate of the critical nucleus generation, being the subject of
the present appendix.

\begin{figure}
  \centering
  \includegraphics[scale=1]{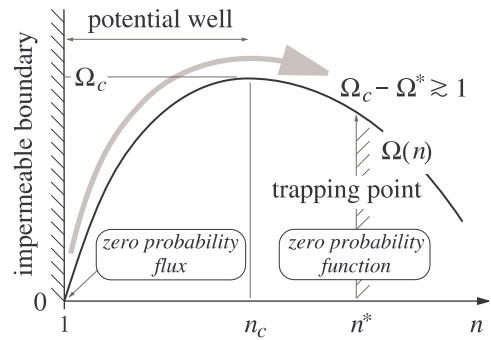}
  \caption{Escaping problem simulating the critical nucleus formation}
  \label{F:pb}
\end{figure}

Mathematically the description of the critical nucleus generation is equivalent
to the problem of a particle escaping form the corresponding potential barrier
(Fig.~\ref{F:pb}). Thereby the rate of the critical nucleus generation, i.e.
the frequency of the traffic breakdown $\nu_\mathrm{bd}$ is represented in
terms of the probability density $\mathcal{F}(t)$ for this particle to escape
from the potential well at a given time $t$ provided initially, $t=0$, it has
been placed near the local minimum (here $n=0$). Namely
\begin{equation}\label{YY}
    \nu_\mathrm{bd} = \mathcal{F}(+0)\,,
\end{equation}
where the value $+0$ of the argument $t$ means that we consider time scales
exceeding substantially the duration of all the transient processes during
which the distribution of the particle inside the potential well attains
locally quasi-equilibrium.

Since the potential relief under consideration is rather special we prefer to
recall briefly the way of deriving the probability $\mathcal{F}(t)$ referring a
reader to the specific literature (see, e.g., book~\cite{Gard}) for details.

The concept of potential well implies that the barrier is sufficiently high,
$\Omega_c\gg 1$, therefore the particle can climb over it due to rare
fluctuations lifting the particle to points at the potential barrier where
$\Omega(n)\gg 1$. If such an event does not lead to escape the particle will
drift back to the the neighborhood of the local minimum $n=0$ whose thickness
is specified by the inequality $\Omega(n)\lesssim 1$. Thereby the subsequent
attempts of escaping may be considered as being mutually independent. After the
particle has climbed over the barrier the force  $-d\Omega(n)/dn$ carries it
away to distant points, making the return impossible. So from this point of
view we may refer to the particle being inside the potential well or having
escaped from it as two its possible states without specifying the particular
position. Therefore the probability $\mathcal{P}(t-t')$ that the particle
remains inside the well at time $t$ if it has being placed in it at time $t'$
obeys the equation:
\begin{equation*}
  \mathcal{P}(t) =
  \mathcal{P}(t-t')\mathcal{P}(t')\quad\mathrm{for}\ 0<t'<t\,.
\end{equation*}
Whence we get the general expression for the function $\mathcal{P}(t)$
\begin{equation*}
  \mathcal{P}(t) = \exp\Big(-\frac{t}{\tau_{\mathrm{life}}} \Big)\,,
\end{equation*}
where $\tau_{\mathrm{life}}$ is a certain constant specified by the particular
properties of a potential well. The latter formula gives us immediately the
general form of the escape probability
\begin{equation}\label{genF}
  \mathcal{F}(t) = -\frac{d\mathcal{P}(t)}{dt} =
  \frac{1}{\tau_{\mathrm{life}}} \exp\Big(-\frac{t}{\tau_{\mathrm{life}}} \Big)\,.
\end{equation}

In order to find the lifetime $\tau_{\mathrm{life}}$ we will deal with the
Laplace transform $\mathcal{F}_L(s)$ of the escape probability $\mathcal{F}(t)$
\begin{equation}\label{genFL}
  \mathcal{F}_L(s):= \int\limits^{\infty}_{0}dt\,\exp(-st)\mathcal{F}(t) =
  \frac{1}{1+s\tau_{\mathrm{life}}}\,,
\end{equation}
whence it follows that in the expansion of $\mathcal{F}_L(s)$ with respect to
$s$ around the point $s=0$
\begin{equation}\label{genFLT}
  \mathcal{F}_L(s) = 1 - s\tau_{\mathrm{life}} + \ldots
\end{equation}
the first order term directly contains the desired lifetime as the coefficient.

Following the standard approach \cite{Gard} we reduce the escaping problem to
finding the first passage time probability. In other words, we assume the
particle never to come back to the potential well if it after climbing the
barrier reaches points where $\Omega_c - \Omega(n) \gtrsim 1$
(Fig.~\ref{F:pb}). The particle may be withdrawn from the consideration or,
what is the same, it will be trapped when reaches for the first time any fixed
point $n^*$ in this region. The time it takes for the particle to reach the
point $n^*$ after overcoming the barrier at the critical point $n_c$ is
ignorable in comparison with the characteristic wait of critical fluctuations.
Thereby the function $\mathcal{F}(t)$ specifies actually the probability of
passing (reaching) the point $n^*$ for the first time at the time moment $t$.
This construction enables us to introduce a more detailed relative function
$\mathcal{F}(n,t)$ giving the probability for the particle initially placed at
the point $0<n<n^*$ to reach first the right boundary $n^*$ of the region under
consideration at the time moment $t$. The left boundary $n=0$ is impermeable
for the particle. Then using the standard technique based on the backward
Fokker-Planck equation conjugated with Eq.~(\ref{e.FP}) we the governing
equation for the function $\mathcal{F}_L(n,s)$
\begin{gather}\label{444}
  \tau_{\infty}s\mathcal{F}_L = \partial^2_n\mathcal{F}_L -
  [\partial_n\Omega(n)][\partial_n\mathcal{F}_L]\\
\intertext{subject to the boundary conditions}
  \mathcal{F}_L(0,s) = \mathcal{F}_L(n^*,s) = 1\,.
\end{gather}
Whence it directly follows that the first order term $\varphi(n)$ in the
expansion of the Laplace transform $\mathcal{F}_L(n,s)$ with respect to $s$
\begin{gather}\nonumber
  \mathcal{F}(n,s) = 1 -s\varphi(n)\,,\\
\intertext{obeys in turn the equation} \label{y1}
  \partial_n^2\varphi(n) -
  [\partial_n\Omega(n)][ \partial_n\varphi(n)] = -\tau_{\infty} \\
\intertext{subject to the boundary conditions} \label{y2}
  \partial_n\varphi(0) = 0
  \quad\mathrm{and}\quad \varphi(n^*) = 0\,.
\end{gather}
The solution of the system~(\ref{y1}) and (\ref{y2}) has the form
\begin{gather}\label{y3}
    \varphi(n)= \tau_{\infty}\int^{n^*}_n dn'\,e^{\Omega(n')}\int^{n'}_0 d n''
    \,e^{-\Omega(n'')} \\
\intertext{and the value $\varphi(0)$ gives us the desired lifetime:}
\label{lt1}
    \tau_{\mathrm{life}} =\varphi(0)\,.
\end{gather}
Inside the potential well the function $\varphi(n)$ takes practically a
constant value mainly contributed by the points $n''$ belonging to the well
bottom, i.e. to the region $\Omega(n'')\lesssim 1$ and by the points $n'$
located near the top of the potential barrier where
$\Omega(n_c)-\Omega(n')\lesssim 1$. This feature leads us immediately to the
approximation
\begin{equation}\label{y4}
    \tau_{\mathrm{life}} \approx \sqrt{2\pi}\tau_{\infty}
    \left[\left|\partial_n^2\Omega(n_c)\right|\right]^{-\tfrac12}
    \left[\partial_n^{\vphantom{2}}\Omega(0)\right]^{-1}\,
    e^{\Omega(n_c)}\,,
\end{equation}
which is the main result of the present appendix.

In particular, for the potential $\Omega(n)$ specified by expression~(\ref{z2})
or (\ref{Om})) formula~(\ref{y4}) gives
\begin{multline}\label{y5}
  \tau_{\mathrm{life}} \approx\sqrt{2\pi n_0}\tau_{\infty}
  \left(\frac{\tau_0}{\tau_{\infty}-\tau_0}\right)^{\tfrac32}\\
  {}\times\left(1-\phi[x_c]\right)^{-1}
  \left(\left|\phi'[x_c]\right|\right)^{-\tfrac{1}{2}}
  e^{\Omega(n_c)}\,.
\end{multline}
Formulae~(\ref{YY}), (\ref{genF}), (\ref{y5}), and (\ref{asa1}) give us
expression~(\ref{e.final}).

\end{document}